\documentstyle[prd,aps,floats]{revtex}
\begin{document}
\draft
%%%%%%%%%%%%%%%%%%%%%%%%%%%%%%%%%%%%%%%%%%%%%%%%%%%%%%%%%%%%%%%%%%%%%%
%
%  Uncomment following four lines and one below for 2 column format
%  and figure insertions.
%
\input epsf
\renewcommand{\topfraction}{0.8}
\twocolumn[\hsize\textwidth\columnwidth\hsize\csname
@twocolumnfalse\endcsname
%%%%%%%%%%%%%%%%%%%%%%%%%%%%%%%%%%%%%%%%%%%%%%%%%%%%%%%%%%%%%%%%%%%%%%
\preprint{CERN-TH/97-27, astro-ph/9702211} 
\title{Open inflation models and gravitational wave anisotropies in 
the CMB}
\author{Juan Garc\'{\i}a-Bellido} 
\address{Theory Division, CERN, CH-1211 Geneva 23, Switzerland} 
\date{May 27, 1997} 
\maketitle
\begin{abstract}
  We study the large scale power spectrum of gravitational wave
  perturbations of the microwave background in the context of
  single-bubble open inflation models. We compute the ratio of tensor
  to scalar contributions to the CMB anisotropies as a function of
  $\Omega_0$, the spectral index $n_S$ and the tunneling parameter
  $2\pi GS_1/H$. We find that gravitational wave anisotropies can be
  very large at small values of this tunneling parameter. We also
  consider the contribution of supercurvature and bubble wall modes
  and find constraints on the parameters of open inflation models from
  the observed temperature anisotropies of the CMB. We show that the
  induced gravity and open hybrid scenarios are compatible with
  present observations for a reasonable range of parameters.
\end{abstract}
\pacs{PACS numbers: 98.80.Cq \hspace{3.8cm} Preprint CERN-TH/97-27,
astro-ph/9702211}

%%%%%%% Comment the next line before submission
\vskip2pc]
%%%%%%%%%%%%%%%%%%%%%%%%%%%%%%%%%%%%%%%%%%%%%%%%%%%%%%%%%%%%%%%%%%%%%%

\section{Introduction}

The inflationary paradigm~\cite{Book} not only gives an explanation
of the large scale homogeneity and isotropy of our observable
universe, but also predicts an almost scale invariant spectrum of
primordial metric perturbations that could be responsible for the
observed temperature anisotropies of the cosmic microwave background
(CMB) as well as the origin of large scale structure~\cite{LL93}.

Until recently, observations of the CMB temperature anisotropies
provided just a few constraints on the parameters of inflationary
models, mainly from the amplitude and the tilt of both scalar and
tensor perturbations' spectrum~\cite{COBE}. Nowadays, with a dozen
experiments looking at different angular scales, we have much more
information about the primordial spectra as well as other cosmological
parameters such as $\Omega_0$, $H_0$, $\Omega_{\rm B}$, etc, see
e.g.~\cite{Charley}. In the near future, high resolution observations
of the microwave background anisotropies with the recently approved
satellites, MAP~\cite{MAP} and PLANCK~\cite{COBRAS}, will determine
the cosmological parameters and the main features of the primordial
spectra of density and gravitational wave perturbations with better
than 1\% accuracy~\cite{JKKS,ZSS,BET}. This means that cosmology is
becoming a phenomenological science, where observations/experiments
determine parameters and allow us to test alternative models of the
universe. A special effort is therefore needed from theoretical
cosmologists in order to predict the essential features as well as
possible variations of the expected CMB power spectrum of temperature
anisotropies. There has been a tremendous progress in this direction
in the last few years and will probably increase as we approach the
time in which the satellites will be launched.

Inflation has generically been associated with a flat universe, due to
its tendency to drive the spatial curvature so effectively to zero.
However, it is now understood that inflation comprises a wider class
of models, some of which may give rise to an open universe at
present~\cite{open,BGT,LM,Green}. Such models generically contain a
field trapped in a false vacuum, which tunnels to its true vacuum via
the nucleation of a single bubble, inside which a second period of
inflation drives the universe to almost flatness. The original
motivation of open inflation as a model where one could reconcile a
large value of the Hubble constant~\cite{HST} with the large estimated
age of globular clusters~\cite{GC} is no longer essential due to the
recent recalibration of distances by the Hipparcos
satellite~\cite{Hipparcos}, which has brought down both the rate of
expansion, $H_0=60\pm10$ km/s/Mpc, and the age of the universe,
$t_0=12\pm2$ Gyr, thus becoming compatible with an Einstein-de Sitter
model, see e.g.~\cite{Dekel}. Large scale structure observations,
however, seem to be in conflict with a flat universe and $h>0.5$,
since they favor low $\Omega_0 h$, see e.g.~\cite{Andrew}. Apart from
still uncertain cosmological observations, open inflation models have
several interesting theoretical features that single them out from
other cosmological models, in particular the way they solve the
homogeneity problem independently from the flatness problem~\cite{LM}.
Furthermore, if future observations determine $\Omega_0$ to be less
than 1 with better than 1\% accuracy, we will have to invoque open
inflation models to explain the large scale homogeneity.

In open inflation models the origin of structure is still related to
amplified quantum fluctuations of the field that drives inflation
inside the bubble~\cite{sasaki,BGT}. A distinct feature of these
models is that in the spectrum of metric perturbations there appears a
discrete supercurvature mode~\cite{LW,GZ}, associated with the open de
Sitter vacuum~\cite{sasaki,YST}, as well as a mode associated with the
bubble wall fluctuations at
tunneling~\cite{LM,Hamazaki,bubble,Garriga}, whose contribution could
be made small in some of the models~\cite{super,induced}. Furthermore,
there is some evidence that the observations made in a wide range of
scales, from horizon size to large clusters of galaxies, constrain
open inflation models (with small $\Omega_0\sim0.3$--$0.4$) to have a
`tilted' spectrum of density perturbations with spectral index
$n_S>1$~\cite{Pedro,WS}, and essentially no other contribution, either
from gravitational waves or supercurvature modes. In order to account
for these observations, we have recently proposed a tilted hybrid
model of open inflation~\cite{GBL} and computed the scalar 
component of the CMB power spectrum. For an alternative way of
producing a large tilt see Ref.~\cite{alter}.

Apart from scalar metric perturbations, open inflation also produces a
primordial spectrum of gravitational waves, whose amplitude and scale
dependence in single-bubble open inflation models has only recently
been known~\cite{TS}. In order to compare with observations one has to
compute the corresponding angular power spectrum $C_l$ of CMB
temperature fluctuations~\cite{HW}. Since the gravitational wave
contribution to the power spectrum decays beyond $l\sim30$, it is
enough to consider the large scale (low multipole) tensor power
spectrum, where gravitational redshift is the dominant effect, without
the need of large computer codes to calculate the full power spectrum.
In this paper we will calculate the first ten multipoles of the CMB
power spectrum for both scalar and tensor components in an open
universe, and determine the ratio of tensor to scalar contributions as
a function of $\Omega_0$ and other model parameters. We will then
constrain general single-bubble open inflation models from such a
tensor component of the CMB anisotropies.

\section{Quantum tunneling and slow-roll inflation}

We will concentrate here in the single-bubble open inflation models
with two fields, the tunneling field $\sigma$ and the inflaton field
$\phi$. The former determines the geometry of the bubble and the
latter produces the inhomogeneities in the metric responsible for the
observed temperature anisotropies of the microwave background.

The $\sigma$ field is initially trapped in its false $(F)$ vacuum and
then tunnels to the true $(T)$ vacuum producing a single bubble. The
extremal instanton action corresponds to the $O(3,1)$ symmetric
bubble~\cite{Parke}
\begin{eqnarray}\label{bounce}
S_B(R) & = & 2\pi^2 R^3 S_1 \nonumber\\%[2mm]
& + & {4\pi^2\over\kappa^2} \Big[ {1\over H_T^2}
\Big((1 - R^2 H_T^2)^{3/2} - 1\Big) \nonumber\\%[2mm]
& - & {1\over H_F^2} \Big((1 - R^2 H_F^2)^{3/2} - 1\Big)\Big] \,,
\end{eqnarray}
where we have taken into account the contributions from the wall
(first term) and the interior of the bubble (in brackets). Here $R$ is
the radius of the bubble, $\kappa^2=8\pi/M_{\rm Pl}^2$, $H_F^2\equiv
\kappa^2 U_F/3$, where $U_F$ is the energy density in the false
vacuum (and similarly for $H_T$ in the true vacuum) and
\begin{equation}\label{S1}
S_1 = \int_{\sigma_F}^{\sigma_T} d\sigma\,[2(U(\sigma)- U_F)]^{1/2}\,.
\end{equation}
For the thin wall approximation to be valid we require that the width
of the bubble wall, $\Delta R$, be much smaller than its radius of
curvature, $\Delta R/R \simeq H_T\,(\Delta\sigma)/[2(U_0 - U_F)]^{1/2}
\ll 1$, where $U_0$ is the value of the potential at the maximum. The
only requirement is that the barrier between $\sigma_F$ and $\sigma_T$
be sufficiently high, i.e. $U_0 \gg \Delta U = U_F-U_T$.
 
The radius of curvature of the bubble wall is that for which the
bounce action (\ref{bounce}) is an extremum. An exact
solution~\cite{Parke} can be written in terms of dimensionless
parameters $a$ and $b$,
\begin{eqnarray}\label{alpha}
R H_T & = & \Big[1+(a+b)^2\Big]^{-1/2} 
\equiv (1 + \Delta^2)^{-1/2}\,,\\[2mm] \label{ab}
a & \equiv & {\Delta U\over3S_1H_T} \,, \hspace{1cm}
b \equiv {\kappa^2S_1\over4H_T} \,.
\end{eqnarray}
Since $S_1\sim U_0/M \sim M(\Delta\sigma)^2$ for a mass $M$ in the
false vacuum, the parameter $a\simeq (\Delta U/U_0)\,M/H_T$, which
characterizes the degeneracy of the vacua, can be made arbitrarily
small by tuning $U_T\simeq U_F$. On the other hand, the parameter
$b\simeq (\Delta\sigma/M_{\rm Pl})^2 M/H_T$, which characterizes the
width of the barrier, is not a tunable parameter and could be very
large or very small depending on the model. It turns out that the
amplitudes of the bubble wall fluctuations and gravitational wave
perturbations of the CMB strongly depend on the value of this
parameter, as we will discuss in Section VI.

\begin{figure}[t]
\centering
\hspace*{-4mm}\leavevmode\epsfysize=5.6cm \epsfbox{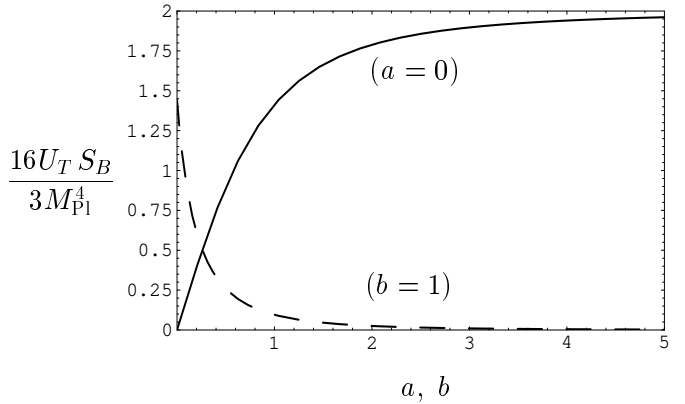}\\
\caption[fig1]{\label{fig7} The bounce action $S_B$ as a function 
  of $b$ for $a=0$ (continuous line) and as a function of $a$ for
  $b=1$ (dashed line), in units of $3M_{\rm Pl}^4/16U_T$. It is clear
  that for a large range of parameters, the bounce action easily
  satisfies $S_B \gg 1$.}
\end{figure}

In order to prevent collisions with other nucleated bubbles (at least
in our past light cone) it is necessary that the probability of
tunneling be sufficiently suppressed. For an open universe of
$\Omega_0 > 0.2$, this is satisfied as long as the bounce action $S_B
> 6$, see Ref.~\cite{open}. This imposes only a very mild constraint
on the tunneling parameters $a$ and $b$, see Fig.~\ref{fig7}, as long
as the energy density in the true vacuum satisfies $U_T \ll M_{\rm
  Pl}^4$.

Just after tunneling the field $\sigma$ is held in its true vacuum 
while the field $\phi$ is free to move down its potential, driving
a short period of inflation responsible for the approximate flatness
of our present universe. The equations of motion of the scalar field
during inflation (where we can soon neglect spatial curvature) are
\begin{eqnarray}
H^2\Big(1-{\epsilon\over3}\Big) &=& 
{\kappa^2\over3}\,V(\phi) \,,\label{H2}\\ 
3H\dot\phi\,\Big(1-{\delta\over3}\Big) &=& -\,V'(\phi) \,,\label{dphi}
\end{eqnarray}
where $\epsilon$ and $\delta$
are defined in terms of the fundamental parameter $H(\phi)$,
\begin{eqnarray}
\epsilon &\equiv& - {\dot H\over H^2} = 
{2\over\kappa^2}\,\Big({H'(\phi)\over H(\phi)}\Big)^2
\,,\label{ep}\\ 
\delta &\equiv& - {\ddot \phi\over H\dot\phi} =
{2\over\kappa^2}\,\Big({H''(\phi)\over H(\phi)}\Big)\,.\label{del}
\end{eqnarray}
Note that for inflation to occur, we need $\epsilon<1$. The
number of $e$-folds of inflation can be computed as
\begin{equation}
N(\phi) = {\kappa^2\over2}\,\int {H d\phi\over H'(\phi)} =
\int {\kappa d\phi\over\sqrt{2\epsilon}}\,.
\end{equation}

In the slow-roll approximation, we have $\epsilon\ll1$ and 
$\delta\simeq \eta-\epsilon\ll1$, where~\cite{LL93}
\begin{eqnarray}
\epsilon &=& {1\over2\kappa^2}\,\Big({V'(\phi)\over V(\phi)}\Big)^2
\ll 1 \,,\label{eps}\\ 
\eta &=& {1\over\kappa^2}\,\Big({V''(\phi)\over V(\phi)}\Big)
\ll 1 \,.\label{eta}
\end{eqnarray}
All quantities of interest can then be expressed in terms of these
slow-roll parameters. In particular, the amplitude and tilt of
the primordial spectrum of density and gravitational wave
perturbations. We will concentrate on those observables in the
following sections.

\section{Scalar and tensor metric perturbations}\label{metric}

We briefly review here the theory of gauge invariant scalar and tensor
metric perturbations. The most general linearly perturbed metric can
be written as~\cite{Bardeen}
\begin{eqnarray}\label{pertbn}
ds^2 & = & a^2(\eta)[- (1+2A) d\eta^2 + 2B_{|i}\,dx^id\eta
\nonumber \\ %[2mm]
& + & \{ (1+2{\cal R}) \gamma_{ij} + 2E_{|ij} + 2h_{ij}\} 
dx^i dx^j ] \,,
\end{eqnarray}
where $\{i,j\}$ label the 3-dimensional open space coordinates with
metric $\gamma_{ij}$. The gauge invariant tensor perturbation $h_{ij}$
corresponds to a transverse traceless gravitational wave, $\nabla^i
h_{ij} = h_i^{\ i} = 0$. The four scalar perturbations are not gauge
independent. Under a gauge transformation $\,\tilde\eta = \eta +
\xi^0(\eta,x^k)$, $\,\tilde x^i = x^i + \gamma^{ij} \xi_{|j}
(\eta,x^k)$, they transform as
\begin{eqnarray}\label{trans1}
\tilde A & = & A - {\xi^0}' - {a'\over a}\xi^0 \,, \hspace{1cm}
\tilde{\cal R} = {\cal R} - {a'\over a}\xi^0 \,,\\ %[2mm]
\label{trans2}
\tilde B & = & B + \xi^0 - \xi' \,, \hspace{1.5cm}
\tilde E = E - \xi \,,\\
&& \hspace{2cm} \tilde h_{ij} = h_{ij} \,,
\end{eqnarray}
where a prime denotes derivative with respect to conformal time
$\eta$. There are however only two gauge invariant gravitational
potentials,
\begin{eqnarray}
\Phi & = & A + {1\over a}\,[a(B-E')]'\,,\label{GI}\\ %[2mm]
\Psi & = & {\cal R} + {a'\over a}\,(B-E')\,,
\end{eqnarray}
which are related through the perturbed Einstein equations,
\begin{eqnarray}\label{EEQ}
\Phi + \Psi & = & 0 \,,\\ %[2mm]
2\,{k^2 - 3K\over a^2}\,\Psi & = & \kappa^2\delta\rho\,.
\end{eqnarray}
Here $\delta\rho$ is the gauge invariant density
perturbation~\cite{Bardeen}.  

In linear perturbation theory, the scalar metric perturbations can be
separated into $A(\eta, x^i) \equiv A(\eta)Q(x^i)$,\footnote{From now
  on $A$, $B$, etc. stand for the $\eta$-dependent functions.} where
$Q(x^i)$ are the scalar harmonics, eigenfunctions of the Laplacian
\begin{equation}\label{Helmholtz}
\nabla^2 Q(r,\theta,\phi) = - k^2\,Q(r,\theta,\phi)\,.
\end{equation}
These solutions have the general form~\cite{Harrison}
\begin{equation}\label{harmonics}
Q_{klm}(r,\theta,\phi) = \Pi_{kl}(r)\, Y_{lm}(\theta,\phi)\,,
\end{equation}
where $Y_{lm}$ are the usual spherical harmonics. 

Furthermore, the gravitational wave perturbations can also be written
as $h_{ij}(\eta, x^k) \equiv h(\eta)\,Q_{ij}(x^k)$, where $Q_{ij}$ are
the transverse traceless tensor harmonics, $k^iQ_{ij} = Q_i^{\ i} =
0$, satisfying the same equation (\ref{Helmholtz}) as the scalar
harmonics~\cite{Bardeen,Harrison}. The radial part of the scalar and
tensor harmonics in an open universe can be found in Appendix~A.

We are interested in the time evolution of these perturbations during
the matter era. The gauge invariant scalar and tensor perturbations
satisfy the following equations during this era,
\begin{eqnarray}
\Phi'' + 3{a'\over a} \Phi' - 2K\,\Phi & = & 0 \,,\label{phi}\\ %[2mm]
h'' + 3{a'\over a} h' + (k^2 + 2K)\,h & = & 0\,,\label{hache}
\end{eqnarray}
where $K=-1\,(0)$ for an open (flat) universe. Metric perturbations
remain constant outside the Hubble scale during inflation and
radiation era and start to evolve as soon as they re-enter during the
matter era, where they create temperature anisotropies on large
scales.

\section{Temperature anisotropies}
 
Quantum fluctuations of the inflaton field $\phi$ during inflation
produce long-wavelength scalar curvature perturbations and tensor
(gravitational wave) perturbations, which leave their imprint in the
CMB anisotropies. Open inflation generates three different types of
scalar modes: those that cross outside during the second stage of
inflation and constitute a continuum of subcurvature
modes~\cite{sasaki,BGT}; a discrete supercurvature mode~\cite{LW},
associated with the open de Sitter vacuum~\cite{YST}, and a mode
associated with the bubble wall fluctuations at
tunneling~\cite{bubble,Garriga,Cohn}, all of which induce temperature
anisotropies in the microwave background. We have already considered
the supercurvature and bubble wall mode in previous
publications~\cite{induced,GBL}. We will concentrate here in the
continuum of scalar and tensor subcurvature modes.

Metric perturbations give rise to temperature fluctuations when
they re-enter the horizon, via gravitational redshift. The dominant
effect on large scales is known as the Sachs-Wolfe effect~\cite{SW67}.
Due to this effect, scalar metric perturbations on the surface of last
scattering are responsible for temperature fluctuations in the
CMB with amplitude given by~\cite{SW67,LL93}
\begin{eqnarray}\label{SWE}
{\delta T\over T}(\theta,\phi) &=& 
{1\over3}\,\Phi(0) \,Q(\eta_0,\theta,\phi) \nonumber\\
&+& 2 \int_0^{\eta_0}\,dr\,\Phi'(r) \,Q(\eta_0-r,\theta,\phi)\,,
\end{eqnarray}
where $\eta_0$ is the present conformal time and $\eta_{LSS} \simeq 0$
corresponds to the last scattering surface. Note that this expression
is valid only for adiabatic fluctuations like those produced by
inflation, see Ref.~\cite{LL93}. The first and second terms are called
the `intrinsic' and `integrated' Sachs-Wolfe effect respectively.

Tensor metric perturbations on the last scattering
surface also create temperature fluctuations with amplitude~\cite{SW67}
\begin{equation}\label{SWT}
{\delta T\over T}(\theta,\phi) = \int_0^{\eta_0}\,dr\,h'(r) \,
Q_{rr}(\eta_0-r,\theta,\phi)\,,
\end{equation}
where $Q_{rr}$ is the $rr$-component of the tensor harmonic along the
line of sight.

Temperature anisotropies in the cosmic microwave background are
usually given in terms of the two-point correlation function or power
spectrum $C_l$, defined by an expansion in multipole number $l$,
\begin{equation}\label{2pc}
\left\langle{\delta T\over T}(\hat n)\cdot{\delta T\over T}
(\hat{n}')\right\rangle_{\hat n\cdot\hat{n}'=\cos\theta} = 
\sum_l {2l+1\over4\pi}\,C_l\,P_l(\cos\theta)\,.
\end{equation}

We are mainly interested in the large scale (low multipole number $l$)
temperature anisotropies since it is there where gravitational waves
could become important. After $l\sim30$, the tensor power spectrum
drops down~\cite{Starobinsky} while the density perturbation spectrum
increases towards the first acoustic peak, see Ref.~\cite{Silk}. On
these large scales the dominant effect is gravitational redshift via
the Sachs-Wolfe effect. Fortunately, this effect can be easily
computed without the need for a CMB code. I will review how this is
done in the case of both flat and open universe power spectra, and
find the relation between the scalar and tensor component of the CMB
anisotropies.

\section{Flat universe power spectra}

In this section we will briefly review the computation of the large
angle (low multipole) scalar and tensor power spectra in a flat
universe. On those scales ($l<30$), the dominant contribution comes
from the Sachs-Wolfe effect~\cite{SW67}. On smaller scales ($l>30$),
the scalar component gets contributions not only from gravitational
perturbations but also from density and velocity fluctuations, which
induce a peak in the spectrum on scales associated with the causal
size of the universe at last scattering~\cite{Mark}. On the other
hand, the tensor component is not coupled to these last sources of
perturbations and decays for large multipole numbers.

\subsection*{Scalar modes}

In a flat universe during the matter era, $a(\eta) \propto \eta^2$,
the growing mode solution of the scalar perturbation equation
(\ref{phi}) is $\Phi = (3/5) {\cal R} =$ constant, where ${\cal R}$ is
the primordial {\em comoving} curvature perturbation during
inflation~\cite{Lyth}. Thus the induced temperature fluctuation on
large scales becomes
\begin{equation}\label{SWS}
{\delta T\over T} = {1\over5}\,{\cal R}\,Q\,,
\end{equation}
where $Q_{klm} = \sqrt{2/\pi}\ k j_l(kr)\,Y_{lm}(\theta,\phi)$ are the
flat scalar harmonics, with radial parts given by spherical Bessel
functions. The corresponding power spectrum (\ref{2pc}) can be written
as
\begin{equation}\label{powerS}
C_l^S = {4\pi\over25}\,\int_0^\infty {dk\over k}\,
{\cal P}_{\cal R}(k)\,j_l^2(k\eta_0)\,,
\end{equation}
where ${\cal P}_{\cal R}(k)$ is the primordial spectrum of scalar
perturbations, defined by
\begin{equation}\label{specS}
\langle{\cal R}_k {\cal R}_{k'}\rangle = {2\pi^2\over k^3}
{\cal P}_{\cal R}(k)\,\delta(k-k')\,.
\end{equation}
It is possible to compute the primordial scalar spectrum during
inflation~\cite{LL93},
\begin{equation}\label{primS}
{\cal P}_{\cal R}(k) = {\kappa^2\over2} \Big({H_T\over2\pi}\Big)^2\,
{1\over\epsilon}\,,
\end{equation}
in the slow-roll approximation. The corresponding tilt of the scalar
spectrum can be defined as
\begin{equation}\label{nS}
n_S - 1 \equiv {d\ln{\cal P}_{\cal R}(k)\over d\ln k} \simeq
- 6\epsilon + 2\eta \,,
\end{equation}
again in the slow-roll approximation, see Eqs.~(\ref{eps},\ref{eta}).
For a scale-invariant spectrum, $n_S = 1$, we can integrate
(\ref{powerS}) to give
\begin{equation}\label{CLS}
l(l+1)\,C_l^S = {2\pi\over25}\,{\cal P}_{\cal R} = {\rm constant}\,.
\end{equation}
As a consequence it is customary to plot the angular power spectrum
as $l(l+1)\,C_l$. On the other hand, for $n\neq1$, the
scalar power spectrum is a complicated function of multipole
number~$l$, see Ref.~\cite{BE87,LL93}
\begin{equation}\label{CLST}
C_l^S = {2\pi\over25}\,{\cal P}_{\cal R}\,{\Gamma({3\over2})\,
\Gamma({3-n\over2})\Gamma(l+{n-1\over2})\over
\Gamma(2-{n\over2})\Gamma(l+2-{n-1\over2})}\,.
\end{equation}
As mentioned above, the Sachs-Wolfe effect is the dominant
contribution only for $l\leq30$, and in fact this last formula breaks
down even at moderate $l\sim20$ due to the rise to the first acoustic
peak, see Ref.~\cite{Bunn}.

\subsection*{Tensor modes}

In a flat universe, the growing mode of the tensor metric perturbation
equation (\ref{hache}) during the matter era is given by $h(\eta) =
h\,G_k(\eta)$, where $h$ is the primordial gravitational wave
perturbation during inflation, which remains constant on large scales
and thus can be related to that at re-entry during the matter era, and
\begin{equation}\label{Gk}
G_k(\eta) = 3\, {j_1(k\eta)\over k\eta}
\end{equation}
is normalized so that $G_k(0)=1$ at the last scattering surface. The
induced temperature fluctuation amplitude (\ref{SWT}) is then
\begin{equation}\label{SWQ}
{\delta T\over T} = \int_0^{\eta_0}\,dr\,h\,G_k'(\eta_0-r)\,
Q_{rr}(r)\,,
\end{equation}
where $Q_{rr}$ is the $rr$-component of the flat universe tensor
harmonic,
\begin{equation}\label{Qrr}
Q_{rr}(r) = \left[{(l-1)l(l+1)(l+2)\over\pi k^2}\right]^{1/2}\,
{j_l(kr)\over r^2}\,.
\end{equation}
The corresponding tensor power spectrum can be written 
as~\cite{AbbottWise,Starobinsky}
\begin{equation}\label{powerT}
C_l^T = {9\pi\over4}\,(l-1)l(l+1)(l+2)\,\int_0^\infty {dk\over k}\,
{\cal P}_g(k)\,I^2_{kl}\,,
\end{equation}
where $I_{kl}$ is given by
\begin{equation}\label{Ikl}
I_{kl} = \int_0^{x_0} dx\,{j_2(x_0-x)\,j_l(x)\over(x_0-x)\,x^2}\,,
\end{equation}
with $x\equiv k\eta$, and ${\cal P}_g(k)$ is the primordial spectrum
of gravitational wave perturbations, defined by
\begin{equation}\label{specT}
\langle h_k h_{k'}\rangle = {2\pi^2\over k^3}
{\cal P}_g(k)\,\delta(k-k')\,.
\end{equation}
It is possible to compute the primordial tensor perturbation
spectrum during inflation,~\cite{Starobinsky,LL93}
\begin{equation}\label{primT}
{\cal P}_{\cal R}(k) = 8\kappa^2 \Big({H_T\over2\pi}\Big)^2\,.
\end{equation}
The corresponding tilt of the gravitational wave spectrum can be
defined as
\begin{equation}\label{nT}
n_T \equiv {d\ln{\cal P}_g(k)\over d\ln k} \simeq
- 2\epsilon \,,
\end{equation}
in the slow-roll approximation (\ref{eps}). For a scale-invariant
spectrum, $n_T = 0$, we can integrate (\ref{powerT}) to
give~\cite{Starobinsky}
\begin{equation}\label{CLT}
l(l+1)\,C_l^T = {\pi\over36}\Big(1 + {48\pi^2\over385}\Big)
\,{\cal P}_g\,A_l\,,
\end{equation}
where $A_l=(1.1184, 0.8789, ...)$ for $l=2,3,...$, which approaches
$A_l=1$ for large multipoles, $l\sim30$, and thus $l(l+1)\,C_l^T$
becomes constant in that limit. Beyond this value of $l$ the
gravitational waves have redshifted away before last scattering and
beyond the limits of the Sachs-Wolfe integral. Thus expression
(\ref{CLT}) is valid only for $l\lesssim30$, see~\cite{Starobinsky}.

We can compute the ratio of tensor to scalar contributions to the CMB
angular power spectrum in a flat universe, using Eqs.~(\ref{CLT}) and
(\ref{CLS}) at large $l$,~\cite{LL93}
\begin{equation}\label{ratio}
R \equiv {C_l^T\over C_l^S} = {25\over72}\Big(1 + 
{48\pi^2\over385}\Big)\,{{\cal P}_g\over{\cal P}_{\cal R}} \simeq
12.4 \, \epsilon \simeq 6.2 \,|n_T|\,,
\end{equation}
in the slow-roll approximation. This is the well known relation
between the ratio $R$ and the tilt of the gravitational wave spectrum.
In most inflationary models the slow-roll parameter $\epsilon$ is so
small that there is essentially no gravitational wave contribution to
the power spectrum and it becomes extremely difficult to measure this
relation, see Ref.~\cite{COBRAS}, unless polarization effects are
taken into account~\cite{SZ}. Therefore, it is worth considering whether
this relation holds in an open universe and how does it characterize
different inflationary models.

\section{Open universe power spectra}

In an open universe, $\Omega<1$, there is a characteristic scale
associated with spatial curvature, $1/a^2 = H^2(1-\Omega)$. As a
consequence the Hubble scale approaches the curvature scale from below
as time progresses, and exceeds it only if there were a non-vanishing
cosmological constant. Furthermore, conformal time $\eta = \int dt/a$
can be interpreted as the coordinate distance to the particle horizon.
Its present value, $\eta_0$, is to very good approximation the
distance to the last scattering surface in units of the curvature
scale and, assuming that the universe is matter dominated since, it is
given by
\begin{equation}\label{eta0}
\eta_0 = \cosh^{-1}\Big({2\over\Omega_0} - 1\Big)\,.
\end{equation}
For $\Omega_0 < 2/(1 + \cosh 1) \simeq 0.786$, the last scattering
surface is located beyond the curvature scale. 

As discussed in the previous section, since the gravitational wave
contribution to the power spectrum decays beyond $l\sim30$, see
Ref.~\cite{HW}, it is enough to consider the large scale (low
multipole) tensor power spectrum, where gravitational redshift is the
dominant effect. In this section we will calculate the first
ten multipoles of the CMB power spectrum for both scalar and tensor
components in an open universe, and determine the ratio of tensor to
scalar contributions as a function of $\Omega_0$ and other model
parameters.

\subsection*{Scalar modes}

The growing mode solution to the scalar perturbation equation
(\ref{phi}) in an open universe during the matter era, with $a(\eta) =
a_0\, (\cosh\eta - 1)$, is $\Phi(\eta) = (3/5){\cal R}\,F(\eta)$,
where ${\cal R}$ is the primordial scalar perturbation after
Hubble-scale crossing during inflation and
\begin{equation}
F(\eta) \equiv 5\,{\sinh^2\eta-3\eta\sinh\eta +4(\cosh\eta-1)
\over(\cosh\eta -1)^3}
\end{equation}
is normalized so that $F(0) = 1$ at last scattering. Due to the
Sachs-Wolfe effect, the induced temperature fluctuations take the 
expression (\ref{SWE}), where $\eta_0$ is given by Eq.~(\ref{eta0}).
The corresponding power spectrum (\ref{2pc}) can be written as
\begin{equation}\label{OpowerS}
C_l^S = {2\pi^2\over25}\,\int_0^\infty {q dq\over 1+q^2}\,
{\cal P}_{\cal R}(q)\,I_{ql}^2\,,
\end{equation}
where $q^2=k^2-1$, and $I_{ql}$ is given by
\begin{equation}\label{Iql}
q I_{ql} = \Pi_{ql}(\eta_0) + 6\int_0^{\eta_0} dr\,\Pi_{ql}(r)\,
F'(\eta_0-r)\,.
\end{equation}
Here ${\cal P}_{\cal R}(q)$ is the primordial spectrum of scalar
metric perturbations, 
\begin{equation}\label{OprimS}
\langle{\cal R}_q {\cal R}_{q'}\rangle = {2\pi^2 
{\cal P}_{\cal R}(q)\over q(1+q^2)}\,\delta(q-q')\,.
\end{equation}

\begin{figure}[t]
\centering
\hspace*{-4mm}\leavevmode\epsfysize=6.8cm \epsfbox{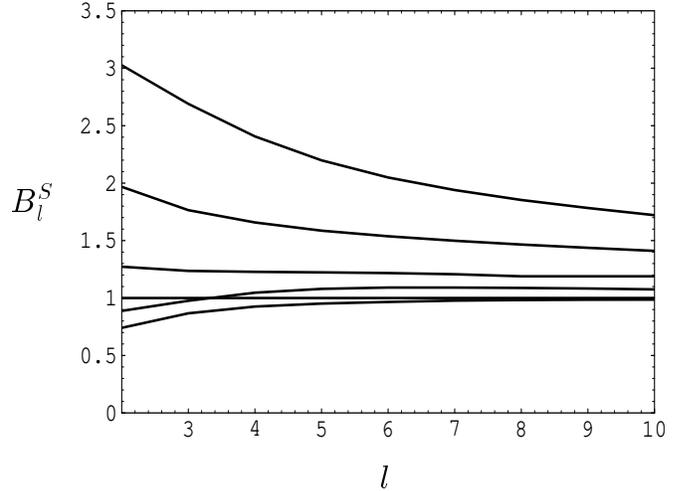}\\
\caption[fig1]{\label{fig1} The first 10 multipoles of the angular power
  spectrum associated with the scale invariant density perturbations,
  $B_l^S$, for $\Omega_0 = 0.3, 0.4, 0.5, 0.6, 0.8$, from top to
  bottom. The straight line at $B_l^S=1$ gives the normalization, for
  $\Omega_0=1$, see Eq.~(\ref{CLO}). }
\end{figure}

\noindent
In the case of single-bubble open inflation models it can be written
as~\cite{YST}
\begin{eqnarray}
{\cal P}_{\cal R}(q) &=& A_S^2 \,f(q)\,, \\ \label{AS2}
A_S^2 &=& {\kappa^2\over2} \Big({H_T\over2\pi}\Big)^2\,
{1\over\epsilon}\,,
\end{eqnarray}
The function $f(q)$ depends on features of the single-bubble open
inflation model, \cite{YST}
\begin{equation}\label{fq}
f(q) = \coth\pi q - {z^2\cos\tilde q + 2q z \sin\tilde q\over
(4q^2 + z^2)\,\sinh\pi q }\,,
\end{equation}
where $\tilde q = q \ln((1+x)/(1-x))$ and
\begin{eqnarray}
x &=& (1-R^2H_T^2)^{1/2} = \Delta\,(1+\Delta^2)^{-1/2}\,, 
\label{x}\\[1mm]
z &=& (1-R^2H_T^2)^{1/2}-(1-R^2H_F^2)^{1/2} \nonumber\\
&=& 2b\,(1+\Delta^2)^{-1/2}\,, \label{z}
\end{eqnarray}
see Eq.~(\ref{alpha}). The function $f(q)$ is linear at small $q$, and
approaches a constant value $f(q) = 1$ at $q\geq2$. We will study its
dependence on the parameters $a$ and $b$ in Appendix B. For scalar
perturbations, the effect of $f(q)$ in the power spectrum $C_l^S$ is
not very important. Therefore, the tilt of the scalar perturbation
spectrum is very approximately given by Eq.~(\ref{nS}).  However, for
a scale invariant spectrum, $n_S=1$, in an open universe, 
$l(l+1)\,C_l^S$ is no longer constant,
\begin{equation}\label{CLO}
l(l+1)\,C_l^S = {2\pi\over25}\,A_S^2\,B_l^S(\Omega_0)\,,
\end{equation}
where $B_l^S$ is a function of $\Omega_0$, which approaches a constant
value at $l\sim20$, where the calculation breaks down as the power
spectrum rises to the first acoustic peak. We have plotted this
function in Fig.~\ref{fig1}, for a scale invariant spectrum and for
various values of $\Omega_0$. For a tilted $n=1.15$ scalar spectrum
see Ref.~\cite{GBL}.

From the four year COBE maps~\cite{COBE}, the overall amplitude and
tilt of the CMB power spectrum has been determined with some
accuracy~\cite{Bond,BLW},
\begin{eqnarray}\label{COBE}
\left[{l(l+1)\,C_l^S\over2\pi}\right]^{1/2} &=&
(1.03 \pm 0.07)\times 10^{-5} \,,\\
n_S &=& 1.02 \pm 0.24\,, 
\end{eqnarray}
assuming that the observed temperature anisotropy on large scales is
{\em solely} determined by the scalar contribution (\ref{CLO}). This
determines the scalar amplitude to be
\begin{equation}\label{conS}
A_S = {1\over\sqrt{\pi\epsilon}}\,{H_T\over M_{\rm Pl}} \simeq
5\times 10^{-5}\,,
\end{equation}
from which we extract a useful relation
\begin{equation}\label{Heps}
H_T \simeq \sqrt\epsilon\,10^{-4}\,{\rm M_{\rm Pl}}\,.
\end{equation}
Note that a tensor contribution $R$ to the total CMB power spectrum
would reduce the amplitude $A_S$ by a factor $(1+R)^{-1/2}$. We will
assume as in Refs.~\cite{Bond,BET} that $R\ll1$. As we will see, in
open hybrid models this may not be a good approximation and we will
have to take it into account.

\subsection*{Tensor modes}

Let us now study the gravitational wave power spectrum in an open
universe. For perturbations that re-enter during the matter era, the
growing mode solution of Eq.~(\ref{hache}) is $h(\eta) = h\,
G_q(\eta)$, where $h$ is the amplitude of the primordial gravitational
wave perturbation, at Hubble-scale crossing during inflation and later on
at re-entry during the matter era, and
\begin{equation}\label{Gq}
G_q(\eta) \equiv 3\,{\sinh\eta\,\sin q\eta-2q\cos q\eta(\cosh\eta-1)
\over q(1+4q^2)\,(\cosh\eta -1)^2}
\end{equation}
is normalized so that $G_q(0) = 1$ at the last scattering surface.
Here $q^2=k^2-3$ for the tensor mode. The induced temperature
fluctuation amplitude is then given by Eq.~(\ref{SWQ}), where $Q_{rr}$
is the $rr$-component of the open universe tensor harmonic,
\begin{equation}\label{OQrr}
Q_{rr}(r) = \left[{(l-1)l(l+1)(l+2)\over2q^2(1+q^2)}\right]^{1/2}\,
{\Pi_{ql}(r)\over\sinh^2r}\,.
\end{equation}
The corresponding tensor power spectrum can be written 
as~\cite{Starobinsky}
\begin{equation}\label{OpowerT}
C_l^T = (l-1)l(l+1)(l+2)\,\int_0^\infty {dq\,\pi^2\,{\cal P}_g(q)\over 
8q^3(1+q^2)^2}\,W^2_{ql}\,,
\end{equation}
where $q^2 = k^2 - 3$ and $W_{ql}$ is given by
\begin{equation}\label{Wql}
W_{ql} = \int_0^{\eta_0} dr\,G_q'(\eta_0-r)\,{\Pi_{ql}(r)\over
\sinh^2r}\,.
\end{equation}
Here ${\cal P}_g(q)$ is the primordial spectrum of open universe
gravitational wave perturbations, defined by
\begin{equation}\label{OspecT}
\langle h_q h_{q'}\rangle = {2\pi^2\,{\cal P}_g(q)\over q(1+q^2)}
\,\delta(q-q')\,.
\end{equation}

\begin{figure}[t]
\centering
\hspace*{-3mm}\leavevmode\epsfysize=6.6cm \epsfbox{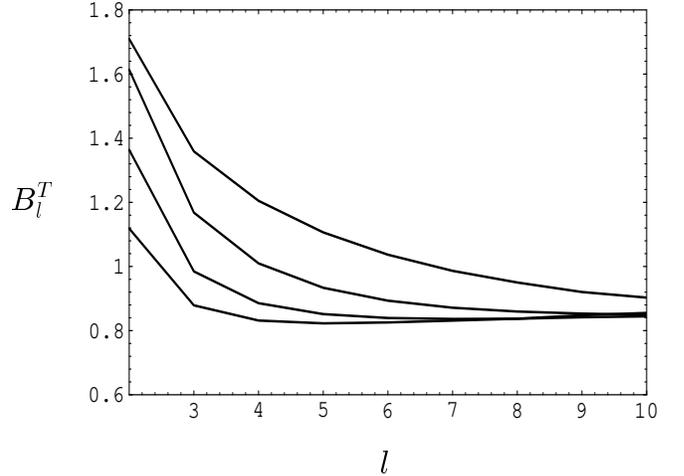}\\
\caption[fig2]{\label{fig2} The first 10 multipoles of the angular power
  spectrum associated with the scale invariant gravitational wave
  perturbations, $B_l^T$, for $\Omega_0 = 0.4, 0.6, 0.8, 1.0$, from
  top to bottom. }
\end{figure}

\noindent
In the case of single-bubble open inflation models it can be written
as~\cite{TS}
\begin{eqnarray}
{\cal P}_g(q) &=& A_T^2 \,f(q) \,, \\
A_T^2 &=& 8\kappa^2 \Big({H_T\over2\pi}\Big)^2\,,
\end{eqnarray}
Here $f(q)$ is the same function (\ref{fq}), which is linear at small
$q$ and thus avoids the infrared divergence at $q=0$ found in
Ref.~\cite{Allen}, but quickly approaches $f(q)=1$ for $q\geq2$. The
effect of $f(q)$ on the tensor power spectrum could become important
only for the first few multipoles. The best situation occurs in the
limit $a\ll1$ and $b\simeq 1$, see Eq.~(\ref{ab}), in which this
function becomes $f(q)\simeq\tanh(\pi q/2)$. However, much larger
contributions are possible for smaller values of $b$, see the
Appendix B.

The corresponding tilt of the gravitational wave spectrum can again be
defined, at large $q$, as in Eq.~(\ref{nT}). For a scale-invariant
spectrum, $n_T = 0$, we can numerically integrate (\ref{OpowerT}) to
give
\begin{equation}\label{OCLT}
l(l+1)\,C_l^T = {\pi\over36}\Big(1 + {48\pi^2\over385}\Big)
\,A_T^2\,B_l^T(\Omega_0)\,,
\end{equation}
where $B_l^T$ is a function of $\Omega_0$, which approaches a constant
value for large multipoles, $l\sim30$, beyond which the calculation
breaks down and the tensor contribution decays very rapidly~\cite{HW}.
We have plotted this function in Fig.~\ref{fig2}, for a scale
invariant spectrum and for various values of $\Omega_0$.

\begin{figure}[t]
\centering
\hspace*{-3mm}\leavevmode\epsfysize=6cm \epsfbox{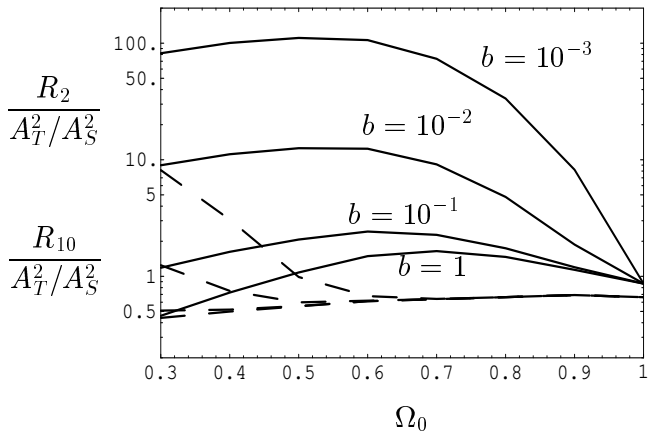}\\
\caption[fig3]{\label{fig3} The ratio of tensor to scalar components
  of the angular power spectrum for the quadrupole (continuous lines),
  $R_2=C_2^T/C_2^S$, and the tenth multipole (dashed lines),
  $R_{10}=C_{10}^T/C_{10}^S$, as a function of $\Omega_0$, for $a=0$
  and $b = 10^{-3}, 10^{-2}, 10^{-1}, 1$, from top to bottom. The
  ratio $R_2$ increases very quickly for very small values of the
  parameter $b$ in the range $\Omega_0\sim 0.3-0.8$, while $R_{10}$
  only grows at small $\Omega_0\lesssim 0.5$.}
\end{figure}

It is now interesting to compute the ratio of tensor to scalar
components of the open universe power spectrum, as a function of
$\Omega_0$. In the limit $l\gg1$, the ratio $R_l = C_l^T/C_l^S$
approaches the flat space limit (\ref{ratio}). However, for small
multipoles, the difference with respect to the $\Omega_0=1$ value
could be large, as we can see from Fig.~\ref{fig3}, where the
corresponding ratio for the quadrupole and the tenth multipole is
shown.\footnote{The peak in the ratio $R_2$ is due to an accidental
  cancellation in the scalar power spectrum between the intrinsic and
  integrated Sachs-Wolfe effect, see Eq.~(\ref{Iql}), which occurs at
  $\Omega_0\simeq0.786$, where the distance to the last scattering
  surface coincides with the curvature scale, i.e. $\eta_0\simeq1$.}
We have also shown the dependence of the ratio $R_2$ with the
tunneling parameter $b$. It is clear from Fig.~\ref{fig3} that values
of $b \lesssim 10^{-2}$ are not allowed, unless $A_T^2/A_S^2 =
16\epsilon$ is very small indeed. However, the ratio $R_{10}$ does not
grow as quickly as $R_2$ and for most values of $\Omega_0$ is below
one.

The ratio $R_2/(A_T^2/A_S^2)=B_2^T(\Omega_0)/B_2^S(\Omega_0)$ depends
very strongly on $b$. For $b=1$ we can approximate it by $1\pm0.5$ in
the range of interest, see Fig.~\ref{fig3}. The condition $R_2 < 1$
then imposes the constraint
\begin{equation}\label{conT}
\epsilon < {1\over16}\,.
\end{equation}
However, for smaller values of $b$, the constraint is much stronger
and also depends on the particular value of $\Omega_0$. For example,
for $b=10^{-2}$ and $\Omega_0\sim 0.3-0.7$, the condition is ten times
stronger, $\epsilon<1/160$, while for $b=10^{-3}$ in the same range,
the condition is a hundred times stronger, $\epsilon<1/1600$. This
means that the flat space relation (\ref{ratio}) between the ratio of
tensor to scalar contributions and the tensor spectral index is no
longer valid. In open inflation models the relation is now a function
of cosmological and model parameters,
\begin{equation}\label{Oratio}
R_l = {C_l^T\over C_l^S} \simeq f_l(\Omega_0,a,b)\,8|n_T|\,
[1-1.3(n_S-1)]\,.
\end{equation}
In the ideal case in which the gravitational wave perturbation can be
disentangled from the scalar component in future precise observations
of the CMB power spectrum, one might be able to test this relation for
a given value of $\Omega_0$. This would then constitute a check on the
tunneling parameter $b$. Such prospects are however very bleak from
measurements of the temperature power spectrum alone, with the next
generation of satellites, see e.g.~\cite{Knox,COBRAS}. At most one can
expect to impose constraints on the parameters of the model from the
absence of a significant gravitational wave contribution to the CMB.
However, taking into account also the polarization power spectrum,
together with the temperature data, one expects to do much better, see
Refs.~\cite{SZ,ZSS} for the case of flat models. Hopefully similar
conclusions can be reached in the context of open models, and CMB
observations may be able to check the generalized consistency relation
(\ref{Oratio}) with some accuracy~\cite{MK}.

\subsection*{Supercurvature mode}

Apart from the continuum of subcurvature modes, in open inflation we
also have a contribution to the microwave background anisotropies
coming from a discrete supercurvature mode, $k^2=0$, which appears in
the spectrum of the inflaton field in open de Sitter when $m_F^2<2
H_F^2$ in the false vacuum~\cite{YST}. The metric perturbation for
this supercurvature mode is~\cite{super,induced}
\begin{equation}\label{RS}
A^2_{SC} = {\kappa^2\over2}\,\Big({H_F^2\over2\pi}\Big)^2\,
{1\over\epsilon} = A^2_S \, {H_F^2\over H_T^2} \,,
\end{equation}
where $A^2_S$ is given by Eq.~(\ref{AS2}).

In Refs.~\cite{induced,GBL}, we computed the corresponding power
spectrum as a function of multipole number $l$. We are only interested
here in the overall contribution of the supercurvature mode to the
total power spectrum, relative to that of the scalar modes. 

\begin{figure}[t]
\centering
\hspace*{-1mm}\leavevmode\epsfysize=6.1cm \epsfbox{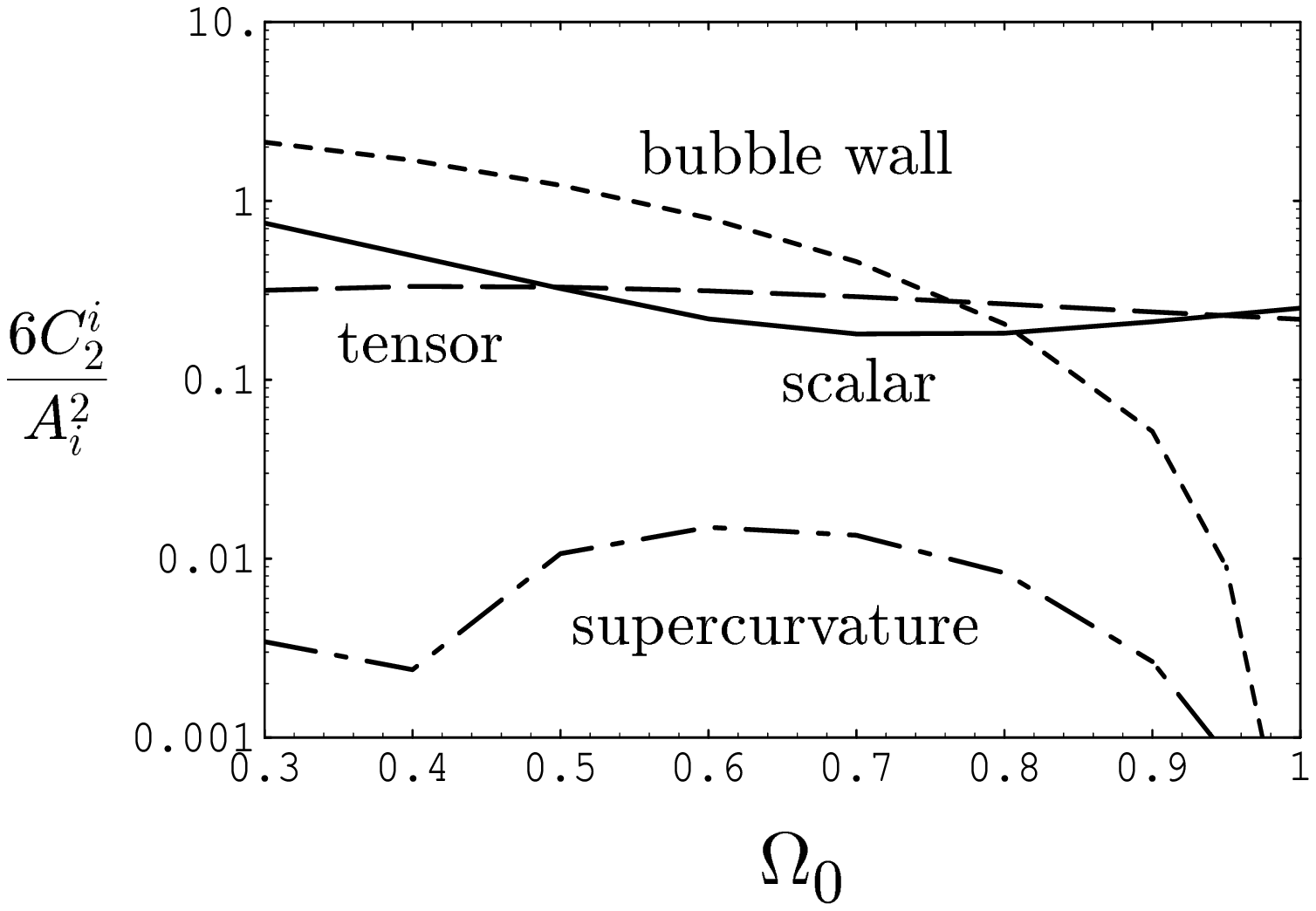}\\ 
\hspace*{-4mm}\leavevmode\epsfysize=6.1cm \epsfbox{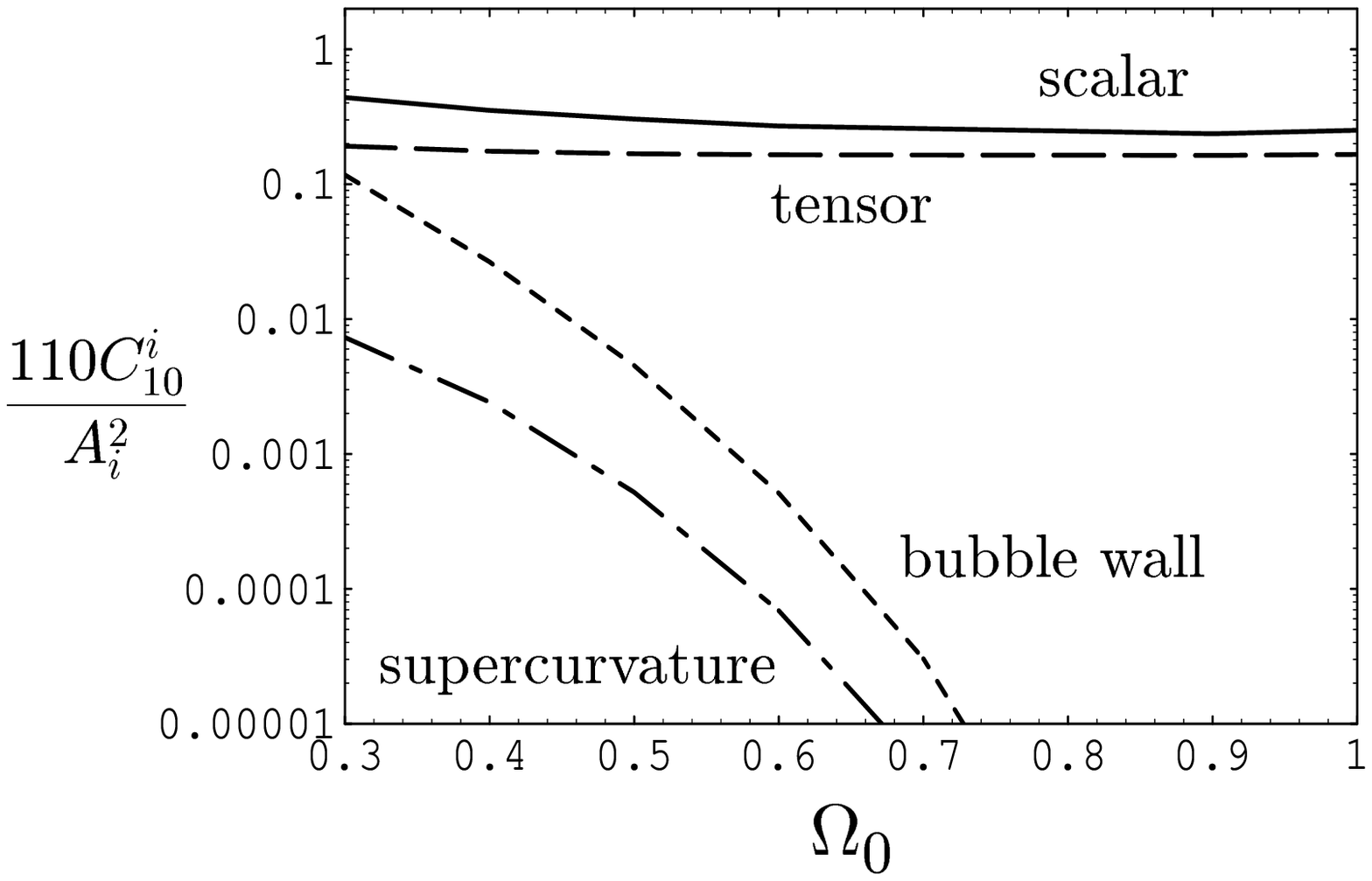}\\
\caption[fig4]{\label{fig4} The quadrupole and tenth multipole
  of the angular power spectrum, normalized to the corresponding
  metric perturbation, for the scalar (continuous lines), tensor
  (dashed lines), bubble wall (dotted lines) and supercurvature modes
  (dotted-dashed lines), as a function of $\Omega_0$. While the
  continuum of scalar and tensor modes do not change much as we
  approach $\Omega_0=1$, the supercurvature and bubble wall modes
  decrease exponentially. }
\end{figure}

The ratio $C_2^{SC}/C_2^S$ is somewhat dependent on $\Omega_0$, see
Fig.~\ref{fig4}, but we can approximate it very roughly by $10^{-2}$
in the range of interest. The condition $C_2^{SC} < C_2^S$ then
imposes a mild constraint on the rates of expansion in the false and
true vacuum,
\begin{equation}\label{conSC}
H_F^2 < 10^2 H_T^2\,.
\end{equation}
For a given value of $\Omega_0$ this constraint can be determined
with greater precision, see Fig.~\ref{fig4}.

\subsection*{Bubble wall mode}

Apart from the continuum of sub\-curvature modes and the dis\-crete
super\-curvature mode, we expect also a contribution from the bubble
wall fluctuations~\cite{bubble,Garriga,Cohn}. These fluctuations
contribute as a transverse traceless curvature perturbation mode with
$k^2=-3$, which never\-theless behaves as a homogeneous random field,
see Refs.~\cite{Hamazaki,modes}. In Ref.~\cite{TS} is was argued that
the bubble wall mode is actually not a discrete mode, once we include
the gravitational backreaction. However, its effect on the CMB
anisotropies can still be computed as if it were a discrete mode
with $k^2=-3$. The constraints on open inflation models from the
absence of this mode's contribution to the CMB do not change much
under this assumption.

The curvature perturbation amplitude for this bubble wall mode can be
computed from~\cite{Garriga,YST}
\begin{equation}\label{wall}
A^2_W = {\kappa^2\over2}\,\Big({H_T^2\over2\pi}\Big)^2\,
{1\over z} = A_S^2\,{\epsilon\over z}\,,
\end{equation}
where $z$ is given in Eq.~(\ref{z}) and $A^2_S$ is the scalar
amplitude of Eq.~(\ref{AS2}).

In Refs.~\cite{induced,GBL}, we computed the corresponding power
spectrum as a function of multipole number $l$. We are only interested
here in the overall contribution of the bubble wall mode to the
total power spectrum, relative to that of the scalar modes. 

The ratio $C_2^W/C_2^S$ is slightly dependent with $\Omega_0$, see
Fig.~\ref{fig4}, but we can approximate it very roughly by 1 in the
range of interest. The condition $C_2^W < C_2^S$ then imposes the
constraint
\begin{equation}\label{conW}
\epsilon < z\,.
\end{equation}
For a given value of $\Omega_0$ this constraint can be determined
with greater precision, see Fig.~\ref{fig4}.

\section{Models of open inflation}

We have obtained generic constraints on open inflation models from the
individual components of the CMB power spectra. It is now necessary to
explore particular models in order to test their viability. In this
section we will concentrate on two concrete open inflation models
which have definite predictions for the complete power spectra of
temperature anisotropies of the CMB: the induced gravity open
inflation model of Ref.~\cite{Green} and the tilted open hybrid
inflation model of Ref.~\cite{GBL}.

\subsection{Induced gravity open inflation}

In Ref.~\cite{induced} we computed only the scalar power spectrum of
CMB anisotropies for the induced gravity open model, without including
the gravitational wave contribution since its primordial spectrum was
not yet known. The tensor primordial spectrum has recently, and
simultaneously, been obtained by several groups, see Refs.~\cite{TS}.
We will show that this model of open inflation is compatible with
CMB observations and contributes with negligible gravitational wave 
perturbations. 

In this model, the tunneling occurs due to the field $\sigma$, with
potential
\begin{equation}\label{US}
U(\sigma) = U_F + {\lambda'\over4}\,\sigma^2 (\sigma-\sigma_0)^2 
- \mu \,U_0 \,(\sigma/\sigma_0)^4\,,
\end{equation}
where $\sigma_0 = M'\sqrt{2/\lambda'}$ corresponds to the true vacuum
and $U_0 = M'^4/16\lambda'$ is the value of the potential at its
maximum, with $\mu\ll1$ for the thin wall approximation to be valid. 

The inflaton field $\varphi$ is non-minimally coupled to gravity, with
coupling $\xi$, and possesses a symmetry breaking potential,
$V(\varphi) = \lambda(\varphi^2-\nu^2)^2/8$, see
Ref.~\cite{Green,induced}. The false vacuum energy density of the
sigma field, $U_F$, determines a stable fixed point for $\varphi$,
in the false vacuum~\cite{induced}
\begin{equation}\label{stat}
\varphi_{\rm st}^2 = \nu^2\Big(1 + {8U_F\over\lambda\nu^4}\Big)
\equiv \nu^2(1+\alpha)\,.
\end{equation}
The corresponding false vacuum rate of expansion in the Einstein frame
is given by $H_F^2 = (\lambda\nu^2/24\xi)\,\alpha/(1+\alpha)$.
After tunneling, the $\sigma$ field lies in its true vacuum at
$U(\sigma_0)\simeq U_F-\mu U_0=0$. In this case, the field $\phi$
is no longer trapped and starts to evolve down its potential, driving
inflation. Immediately after tunneling, the rate of expansion in the
true vacuum, inside the bubble, is related to that in the false vacuum
by~\cite{induced}
\begin{equation}\label{HFT}
H_T^2 = H_F^2\,{\alpha\over1+\alpha}\,.
\end{equation}
Therefore, for parameter $\alpha\geq1$, the rates of expansion are
very similar. This will suppress the contribution of the supercurvature
mode to the CMB power spectrum, see Eq.~(\ref{conSC}).

The amplitude and the tilt of the scalar perturbation spectrum are
determined by the slow-roll parameters soon after tunneling,
\begin{eqnarray}
\epsilon &=& {8\xi\over1+6\xi}\,{1\over\alpha^2}\,,\label{epa}\\
\eta &=& {8\xi\over1+6\xi}\,{1-\alpha\over\alpha^2}\,.\label{et}
\end{eqnarray}
The spectral tilt of scalar and tensor perturbations is always
negative,~\cite{induced}
\begin{eqnarray}
n_S - 1 &=& -\,{8\xi\over1+6\xi}\,{2(2+\alpha)\over\alpha^2}\,,
\label{indnS}\\
n_T &=& -\,{8\xi\over1+6\xi}\,{2\over\alpha^2}\,.\label{indnT}
\end{eqnarray}

Let us now study the contribution from fluctuations of the bubble wall,
Eq.~(\ref{wall}). For that purpose we have to compute the tunneling
parameters $a$ and $b$. From Eqs.~(\ref{ab}) and (\ref{HFT}) we see 
that $ab=1/4\alpha$, while $b$ is given by  
\begin{equation}\label{b2}
b = {\kappa^2 S_1\over4H_T} = {2\pi\over3\lambda'}\,{M'\over H_T}
\Big({M'\over M_{\rm Pl}}\Big)^2\,,
\end{equation}
where $S_1=M'^3/3\lambda'$ is the contribution to the bounce action
coming from the bubble wall, see Eq.~(\ref{S1}), and we have used the
COBE normalization (\ref{Heps}). We will consider here large values of
$M'/H_T$ and thus prevent a supercurvature mode of the $\sigma$ field,
see Ref.~\cite{sasaki}. In this limit, $a\ll1$ and
\begin{equation}\label{zb}
z\simeq {2b\over(1+b^2)^{1/2}}\,.
\end{equation}
The condition (\ref{conW}) on the bubble wall contribution to the CMB 
then determines
\begin{equation}\label{abeps}
b > \epsilon/2\,,
\end{equation}
which is not very difficult to satisfy in the models of 
Ref.~\cite{induced}, where $\epsilon$ is small.

Furthermore, in the limit $a\ll1$, the smallest contribution of the
tensor modes to the CMB anisotropies comes from the parameter
$b\simeq1$, where $f(q) \simeq \tanh(\pi q/2)$, see Appendix B.
Induced gravity models determine the parameter $ab=1/4\alpha$, and it
is always possible to choose $a$ so that $b\simeq1$. In this case, the
constraint on the tensor amplitude comes from the quadrupole, which
requires $R_2(\Omega_0)<1$, see Fig.~\ref{fig3}. For most $\Omega_0$
in the range of interest, this constraint is satisfied provided
$\epsilon\lesssim 1/16$. Again, this is not difficult to satisfy in
the induced gravity open inflation model~\cite{Green,induced}.

For example, in the case $\xi\ll1$, we can have $8\xi = 1/200$,
$\alpha=1$, together with $a=1/4, b=1, \lambda'=0.1$ and
$M' = 10^3\,H_T = 7\times 10^{-3} M_{\rm Pl}$. This gives
$\epsilon = 1/200$, $H_F^2/H_T^2 = 2$, and $z\simeq1.25$, which
satisfies all constraints.

On the other hand, in the case $\xi\gg1$, we have $\alpha=85$,
together with $a=1/340, b=1, \lambda'=0.1$ and $M' = 3\times10^3\,H_T
= 4\times 10^{-3} M_{\rm Pl}$. This gives $\epsilon = 2\times
10^{-4}$, $H_F^2/H_T^2 \simeq 1$, and $z\simeq1.41$, which again
satisfies all constraints.

Induced gravity models are thus viable scenarios of open inflation,
with small and negative spectral tilt and negligible contribution
of gravitational waves to the CMB power spectrum.

\subsection{Open hybrid inflation}

Open hybrid inflation~\cite{GBL} was proposed recently in an attempt
to produce a significantly tilted scalar spectrum in the context of
open models. It is based on the hybrid inflation
scenario~\cite{hybrid}, which has recently received some attention
from the point of view of particle physics~\cite{Guth,GBLW}, together
with a tunneling field which sets the initial conditions inside the
bubble.

In this model there are three fields: the tunneling field $\sigma$,
the inflaton field $\phi$ and the triggering field $\psi$. Quantum
tunneling occurs in the $\sigma$ field due to its coupling to the
$\phi$ field. When $\phi$ drops below a certain value, $\phi_*$, the
true vacuum appears and there is an increasing probability that the
$\sigma$ field will tunnel to it, creating a single bubble inside
which the $\phi$ field will slow-roll down its potential driving
inflation and producing the observed metric perturbations, until it
drops below another scale, $\phi_c$, for which the triggering field
$\psi$ acquires a negative mass and suffers a sudden phase transition
which ends inflation. This second period is known as hybrid inflation,
see Ref.~\cite{hybrid}.

The complete potential is the sum of the open model plus the
hybrid model,~\cite{GBL}
\begin{eqnarray}\label{tunnel1}
 V(\sigma,\phi,\psi) &=& -{M^2 \psi^2\over 2}+
 {\lambda\psi^4 \over 4 } + {M^4\over 4\lambda} \,
 \exp\Big({4\pi\alpha\phi^2\over3M_{\rm Pl}^2}\Big) +
\nonumber \\
&+ & {M'^2 \sigma^2\over2} - \sqrt{\lambda'} M'\sigma^3 +
{ \lambda'\sigma^4\over4}\nonumber \\
 &+& { \phi^2\over2}(g^2\psi^2 + h^2\sigma^2) + \tilde V_0\,,
\end{eqnarray}
where $\tilde V_0\simeq M'^4/\lambda'$ has been added to make the
effective potential vanish in the global minimum, $\phi=0,
\psi=M/\sqrt\lambda, \sigma=(3+\sqrt5)\,M'/2\sqrt{\lambda'}$. Quantum
tunneling occurs at $\phi=\phi_*=M'/h$ and the phase transition that
triggers the end of inflation occurs at $\phi=\phi_c= M/g$.

This model is rather constrained by the potential of the tunneling
field, due to its coupling to the inflaton field. In particular, the
vacuum energy of the $\sigma$ field cannot be much larger than that of
the $\psi$ field, and its contribution to the {\it effective} mass
of the $\phi$ field should also be suppressed,
\begin{eqnarray}\label{Mass}
{h^2 M'^2\over\lambda'} &\leq&
{2\pi\alpha M^4\over3\lambda M_{\rm Pl}^2} \,, \\ \label{FV}
{M'^4\over\lambda'} &\leq& {M^4\over\lambda } \,.
\end{eqnarray}
These constraints might prove to be too strong once we include the
gravitational wave CMB power spectrum.

The rate of expansion during inflation is dominated by the false
vacuum energy of the $\psi$ field,
\begin{equation}
H^2 = {2\pi\over3\lambda}\,{M^4\over M_{\rm Pl}^2}\,.
\end{equation}
The number of $e$-folds is 
\begin{equation}\label{efolds}
N={3\over\alpha}\,\ln\Big({gM'\over hM}\Big) = 55\,.
\end{equation}
The slow-roll parameters become 
\begin{eqnarray}
\epsilon &=& {4\pi\alpha^2\phi^2\over9M_{\rm Pl}^2} \label{epsOHI} \\
\eta &=& {\alpha\over3}\Big(1 + {8\pi\alpha\phi^2\over
3M_{\rm Pl}^2}\Big) \label{etaOHI}
\end{eqnarray}
from which the tilt of the scalar and tensor spectra can be computed as
\begin{eqnarray}
n_S - 1 &=& {2\alpha\over3} - 2\epsilon \,, \label{tiltS} \\
n_T &=& - 2\epsilon \,. \label{tiltT}
\end{eqnarray}

On the other hand, the bubble parameter (\ref{S1}) is here
$S_1=2\sqrt2\,M'^3/3\lambda'$, which substituted into Eq.~(\ref{ab})
gives
\begin{equation}\label{abOHI}
b = 2\sqrt2\,{H\over M'}\,\Big({\lambda\,M'^4\over\lambda'
M^4}\Big)\,.
\end{equation}
This parameter must always be approximately less than one in our
model, since $M'\geq H$ in order to prevent Hawking-Moss
tunneling~\cite{LM} and/or a large supercurvature mode
perturbation~\cite{sasaki}, and $M'^4/\lambda' \leq M^4/\lambda$ for
$\psi$-vacuum domination during inflation. This means that
gravitational waves could give an important contribution to the CMB
anisotropies in these models.

Consider, for instance, the particular values $\alpha=0.25, \, g=0.2,
\lambda=0.1$ and $M=1.5\times10^{-3}M_{\rm Pl}$ for the tilted hybrid
model, and $h=10^{-3}, \lambda'=0.01$ and $M'=M/2$ for the tunneling
field. In that case, Eqs.~(\ref{tiltS})--(\ref{abOHI}) give 
\begin{eqnarray}
n_S &\simeq& 1.15 \,, \hspace{1cm} n_T \simeq - 0.02 \,,\\
b &\simeq& 0.03 \,, \hspace{1.3cm} \epsilon \simeq 0.01 \,.
\end{eqnarray}
Note that the spectral tilts, $n_S$ and $n_T$, are significantly
different from their `canonical' values $n_S=1, n_T=0$.  This is a
generic feature of these models. We also have a large contribution to
the power spectrum coming from the bubble wall fluctuations, see
Eq.~(\ref{conW}), as well as from tensor perturbations, $R_2 \simeq
0.65$ (for small $\Omega_0$, see Fig.~\ref{fig3}), which could be
potentially dangerous. However, the quadrupole is by far the largest
multipole and could be hidden in the cosmic variance uncertainty for
small multipoles~\cite{JKKS}
\begin{equation}\label{cosmic}
\sigma_l = \Big[{2\over(2l+1)f_{\rm sky}}\Big]^{1/2}\, C_l\,,
\end{equation}
where $f_{\rm sky}$ is the fraction of the sky covered by the
particular experiment. For the quadrupole $l=2$ and a typical fraction
$f_{\rm sky}=1/3$, we find that the estimated error $\sigma_2$ is of
the same order as $C_2$ and thus a tensor contribution with $R_2<1$
could be hidden in the CMB temperature maps.  On the other hand, by
the time we reach the tenth multipole, the ratio $R_{10}$ has
decreased considerably, see Fig.~\ref{fig3}, and therefore these
parameters are still allowed by observations.

Open hybrid models are thus viable models of inflation with the
special property of producing a positively tilted spectrum of density
perturbations, which might help the agreement with observations
of large scale structure~\cite{Pedro} and CMB anisotropies~\cite{WS}.

\section{Conclusions}

In the near future, observations of the microwave background will
determine with better than 1\% accuracy whether we live in an open
universe or not. It is therefore crucial to know whether inflation
can be made compatible with such a universe. Single-bubble open 
inflation models provide a natural scenario for understanding the
large scale homogeneity and isotropy. Furthermore, inflationary
models generically predict a nearly scale invariant spectrum of
density and gravitational wave perturbations, which could be
responsible for the observed CMB temperature anisotropies.
Future observations could then determine whether inflationary
models are compatible with the observed features of the CMB power
spectrum. For that purpose it is necessary to know the predicted
power spectrum from inflation with great accuracy. Open models
have a more complicated primordial spectrum of perturbations,
with extra discrete modes and possibly large tensor anisotropies.
In order to constrain those models we have to compute the full
spectrum for a large range of parameters.

In this paper we have computed the large scale angular power spectrum
of temperature fluctuations in the CMB induced by gravitational wave
perturbations in the context of the single-bubble open inflation
models. We have then studied the dependence of the ratio $R$ of tensor
to scalar components with the value of $\Omega_0$ and the tunneling
parameter $2\pi GS_1/H$. We have shown that $R$ increases very quickly
for very small values of this parameter. The flat-space consistency
relation between the ratio $R$ and the tensor spectral index $n_T$ is
now a more complicated relation, which mainly depends on the tunneling
parameter $2\pi GS_1/H$. In the ideal case in which the gravitational
wave perturbation can be disentangled from the scalar component in
future precise observations of the CMB power spectrum, one might be
able to test this relation for a given value of $\Omega_0$. Such
prospects are however very bleak from measurements of the temperature
power spectrum alone, with the recently approved new generation of
satellites, see e.g.~\cite{Knox,JKKS}. At most one can expect to
impose constraints on the parameters of the model from the absence of
a significant gravitational wave contribution to the CMB. However,
taking also into account the CMB polarization power spectrum, together
with the temperature power spectrum, one expects to do much better,
see Refs.~\cite{SZ,ZSS} for the case of flat models. Hopefully similar
conclusions can be reached in the context of open models, and CMB
observations may be able to test the generalized consistency relation
with some accuracy~\cite{MK}.

We have found a set of constraints from scalar, tensor, supercurvature
and bubble wall modes' contribution to the CMB anisotropies that the
parameters of a general open inflation model should satisfy in order
to agree with observations. We have applied such constraints to the
induced gravity and open hybrid inflation models and found a range
of parameters which make them compatible with present observations. In
the future we might be able to determine these parameters with
greater precision.

\section*{Note added}

While writing this paper, we noticed the work of Sasaki et
al.~\cite{new} in the {\tt astro-ph} archive, where similar
conclusions were reached.

\section*{Acknowledgements}

It is a pleasure to thank Andrei Linde for constant encouragement and
in particular for his comments on the best parameters for open hybrid
models.

%%%%%%%%%%%%%%%%%%%%%%%%%%%%%%%%%%%%%%%%%%%%%%%%%%%%%%%%%%%%%%%%%

\appendix

\section{Open universe mode functions}

The open universe scalar mode functions are discussed in
Refs.~\cite{Harrison,modes}. The correctly normalized subcurvature 
scalar modes can be written as
\begin{equation}
\label{PQL}
\Pi_{ql}(r) = N_{ql}\,\tilde\Pi_{ql}(r)\,,
\end{equation}
with
\begin{equation}
N_{ql} = \sqrt{\frac{2}{\pi}} \prod_{n=1}^l (n^2+q^2)^{-1/2} \,,
        \hspace{5mm} N_{q0} = \sqrt{\frac{2}{\pi}} \,,
\end{equation}
where the unnormalized modes $\tilde\Pi_{ql}(r)$ can be generated from the 
first two
\begin{eqnarray}
\label{modesC}
\tilde\Pi_{q0}(r) &=& {\sin qr\over\sinh r}\,,\\[1mm]
\tilde\Pi_{q1}(r) &=& {\coth r\,\sin qr - q\cos qr\over\sinh r}\,,
\end{eqnarray}
through the recurrence relation
\begin{eqnarray}
\label{recurC}
\tilde\Pi_{ql}(r) &=& (2l-1)\,\coth r \,\tilde\Pi_{q,l-1}(r) \nonumber
        \\[1mm]
&& \ -\, [(l-1)^2+q^2]\,\tilde\Pi_{q,l-2}(r)\,.
\end{eqnarray}

In the limit $\Omega_0 \to 1$, the scale factor becomes $a\propto
\sinh^2\eta\to\eta^2$; the eigenvalues of the Laplacian $q^2\to k^2$,
and the scalar eigenfunctions become
\begin{equation}
\Pi_{ql}(r) \to \sqrt{2\over\pi}\,k\,j_l(kr)\,.
\end{equation}

\begin{figure}[t]
\centering
\hspace*{3mm}\leavevmode\epsfysize=6.0cm \epsfbox{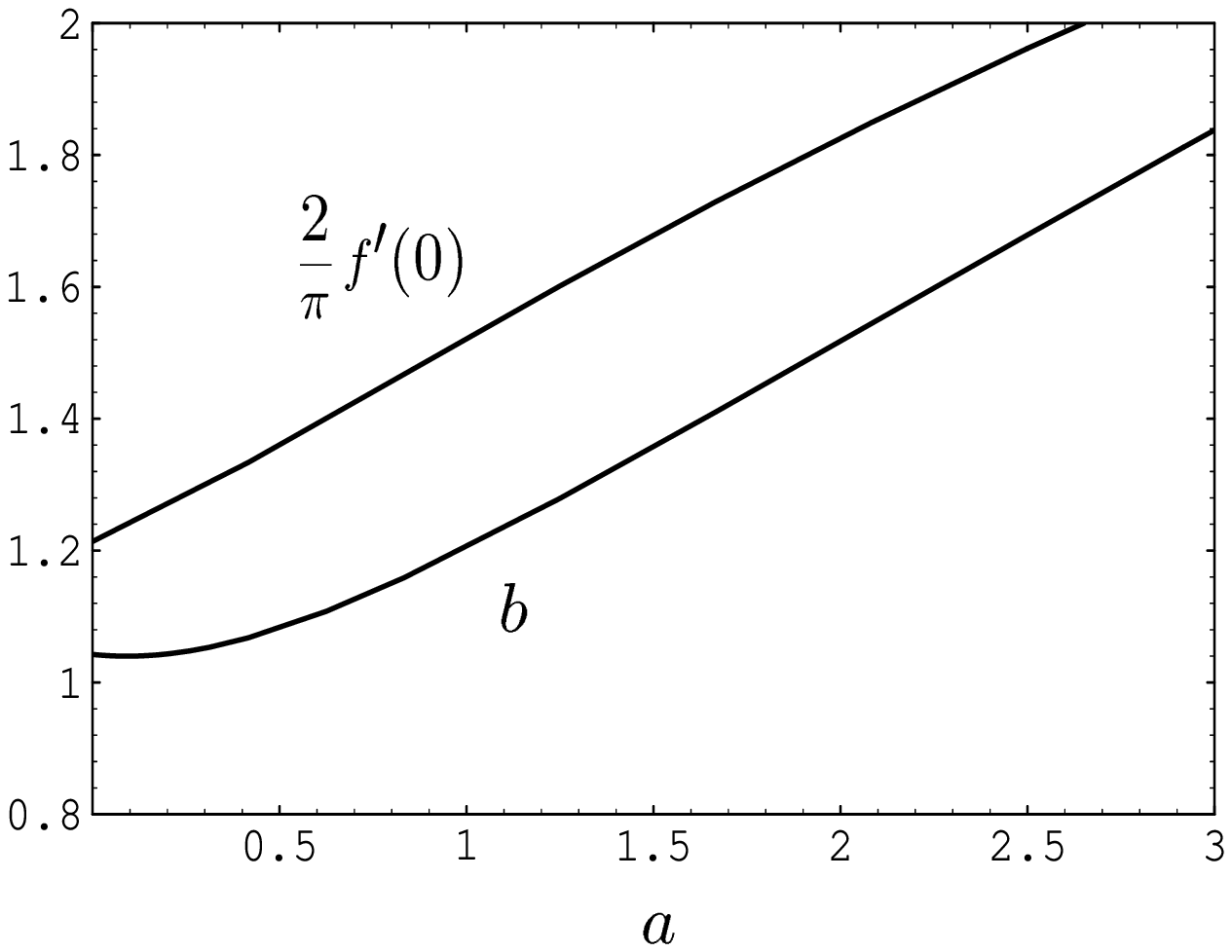}\\
\hspace*{-4mm}\leavevmode\epsfysize=5.8cm \epsfbox{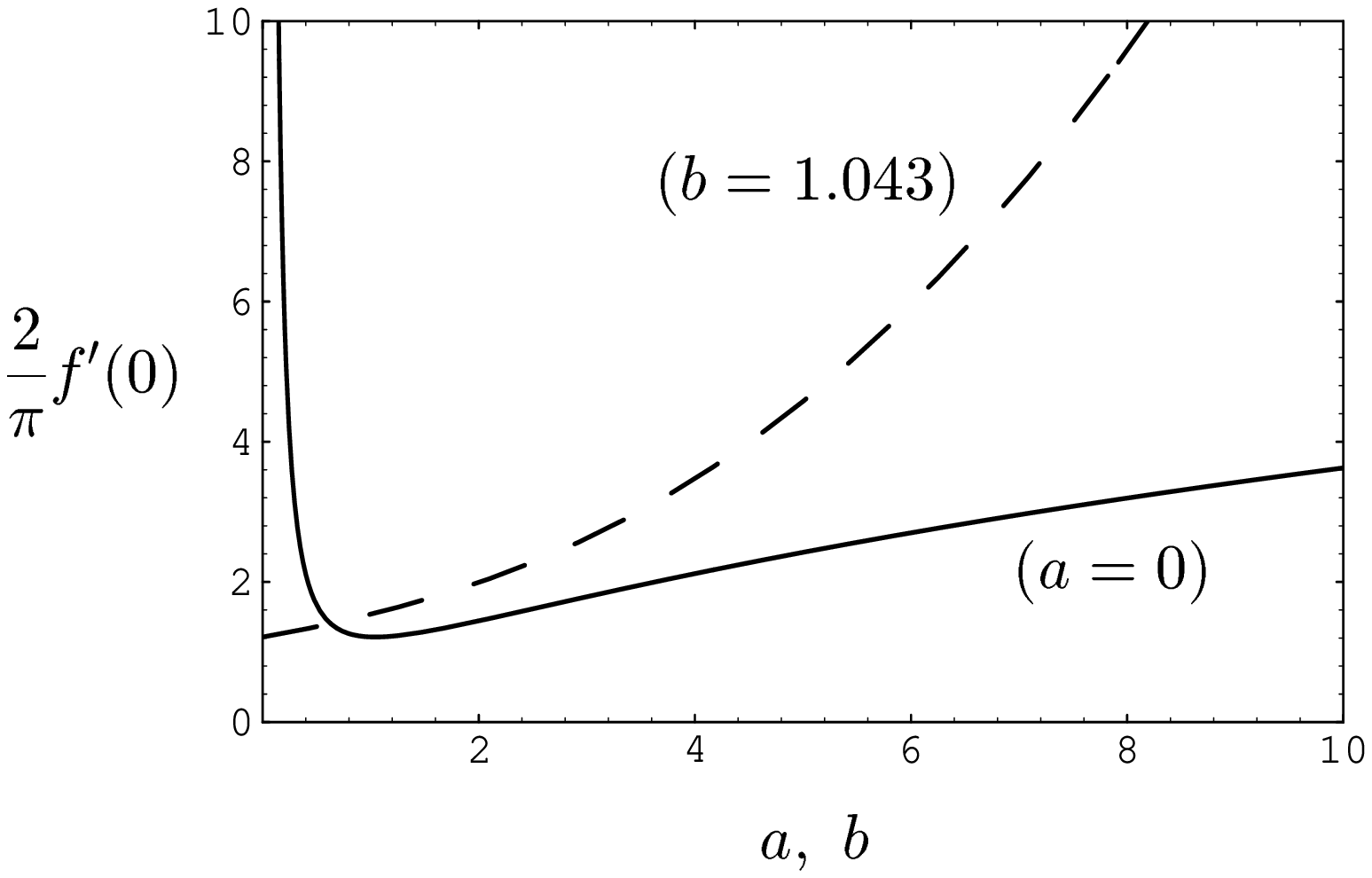}\\
\caption[fig5]{\label{fig5} The top panel shows the minimum slope of 
  $f(q)$ and the corresponding value of $b$ as a function of the
  parameter $a$. They approach asymptotically the values $f'(0) =
  1.214\,\pi/2$ and $b=1.043$, respectively. The lower panel shows the
  sharp increase in the slope of $f(q)$ at small values of the
  parameter $b$, for $a=0$ (continuous line) and at large values of
  $a$, for $b=1.043$ (dashed line). }
\end{figure}

On the other hand, the correctly normalized radial component of the 
tensor modes can be written as~\cite{Harrison}
\begin{equation}
Q_{rr}(r) = \left[{(l-1)l(l+1)(l+2)\over2q^2(1+q^2)}\right]^{1/2}\,
{\Pi_{ql}(r)\over\sinh^2r}\,,
\end{equation}
where $\Pi_{ql}(r)$ is the scalar mode (\ref{PQL}). In the limit
$\Omega_0\to1$, this tensor mode becomes $Q_{rr}$ in Eq.~(\ref{Qrr}), 
as expected, while $G_q(r) \to G_k(r)$.

\section{Dependence of the gravitational wave spectrum on the
  tunneling parameters}

\begin{figure}[t]
\centering\hspace*{-4mm}
\leavevmode\epsfysize=5.8cm \epsfbox{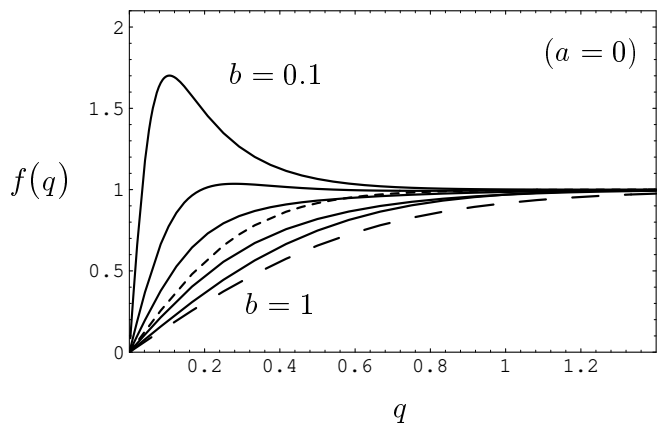}\\
\caption[fig6]{\label{fig6} The spectral function $f(q)$ for $a=0$
  and various values of $b=0.1, 0.2, 0.3, 0.5, 1.0$, from top to
  bottom (continuous lines). As we decrease $b$ the function develops
  a bump at small $q$ which becomes very pronounced for very small
  values of $b$, increasing significantly the gravitational wave
  contribution to the CMB power spectrum. The dashed line corresponds
  to $f(q)=\tanh\pi q/2$ and the dotted line to $f(q)=\tanh\pi q$.}
\end{figure}

In this Appendix we will study the behavior of the function $f(q)$,
Eq.~(\ref{fq}), in the spectrum of gravitational wave perturbations.
This function behaves like $f(q) \simeq \tanh(\pi q/2)$ in a certain
well defined limit. However, for arbitrary values of the tunneling
parameters (\ref{ab}), it can take a very different shape. For 
certain parameters, it increases very quickly at small $q$, thus 
giving a large contribution to the gravitational wave power spectrum
of CMB anisotropies. It is therefore important to study its behaviour
at small $q$, where $f(q)$ is linear. The slope at the origin is
given by
\begin{eqnarray}\label{slope}
f'(0) &=& {\pi\over2} \times \\
&&\hspace{-1.3cm}\nonumber
\left[1 + {1+\Delta^2 + [2b\,\ln(\Delta + \sqrt{1+\Delta^2}) - 
\sqrt{1+\Delta^2}]^2\over\pi^2b^2}\right]\,,
\end{eqnarray}
where $\Delta$ is defined in Eq.~(\ref{alpha}). For $b\simeq1$ we
recover the limiting function, $f(q) \simeq \tanh(\pi q/2)$. For all
other values, the slope (\ref{slope}) is larger and so is the
contribution of the tensor modes to the CMB anisotropies. The best
situation corresponds to those values of $a$ and $b$ for which
(\ref{slope}) is as small as possible.

Let us now study the behaviour of the slope (\ref{slope}) with various
tunneling parameters. In Fig.~\ref{fig5} we show the minimum possible
slope and the corresponding value of $b$ for a given value of $a$. One
can see that the smallest contribution to the CMB comes from $a=0$,
and $b=1.043\simeq1$, for which the slope becomes $1.214\,\pi/2$. This
gives a slightly larger contribution than the minimum function
$\tanh(\pi q/2)$.  On the other hand, for $b<1$, the minimum slope
increases very quickly, as we can see in Fig.~\ref{fig5}. 

We have shown various functions $f(q)$ in Fig.~\ref{fig6}, for $a=0$
and $b=0.1-1$. The effect of a large $f(q)$ at small $q$ is an
increased tensor contribution to the CMB. This effect is very
important for small $\Omega_0$, as we can see in Fig.~\ref{fig3}. In
Ref.~\cite{HW} this large effect at small values of the parameter $b$
was not realized since they assumed that the function $f(q)$ was
bounded between $\tanh(\pi q)$ and $\tanh(\pi q/2)$, what they called
{\it maximal} and {\it minimal} tensor anisotropies. As one can see
from Fig.~\ref{fig6}, larger gravitational wave anisotropies are
possible in certain classes of models.

%%%%%%%%%%%%%%%%%%%%%%%%%%%%%%%%%%%%%%%%%%%%%%%%%%%%%%%%%%%%%%%%%%%

\end{document}